## Research Article

# Emotional Video to Audio Transformation Using Deep Recurrent Neural Networks and a Neuro-Fuzzy System

**Gwenaelle Cunha Sergio** and **Minho Lee**

*School of Electronics Engineering, Kyungpook National University, Daegu 41566, Republic of Korea*

Correspondence should be addressed to Minho Lee; mholee@gmail.com

Generating music with emotion similar to that of an input video is a very relevant issue nowadays. Video content creators and automatic movie directors benefit from maintaining their viewers engaged, which can be facilitated by producing novel material eliciting stronger emotions in them. Moreover, there is currently a demand for more empathetic computers to aid humans in applications such as augmenting the perception ability of visually- and/or hearing-impaired people. Current approaches overlook the video's emotional characteristics in the music generation step, only consider static images instead of videos, are unable to generate novel music, and require a high level of human effort and skills. In this study, we propose a novel hybrid deep neural network that uses an Adaptive Neuro-Fuzzy Inference System to predict a video's emotion from its visual features and a deep Long Short-Term Memory Recurrent Neural Network to generate its corresponding audio signals with similar emotional inkling. The former is able to appropriately model emotions due to its fuzzy properties, and the latter is able to model data with dynamic time properties well due to the availability of the previous hidden state information. The novelty of our proposed method lies in the extraction of visual emotional features in order to transform them into audio signals with corresponding emotional aspects for users. Quantitative experiments show low mean absolute errors of 0.217 and 0.255 in the Lindsey and DEAP datasets, respectively, and similar global features in the spectrograms. This indicates that our model is able to appropriately perform domain transformation between visual and audio features. Based on experimental results, our model can effectively generate an audio that matches the scene eliciting a similar emotion from the viewer in both datasets, and music generated by our model is also chosen more often (code available online at https://github.com/gcunhase/Emotional-Video-to-Audio-with-ANFIS-DeepRNN).

## 1. Introduction

The acquisition and understanding of art is an intrinsic characteristic of humans, and it is one of the main attributes that separates us from other species in the animal kingdom [1]. Zaidel [2] states that art and cognition are deeply intertwined and defines art as being a human activity correlated with abstract and symbolic cognition. This ability to learn to use symbols and understand abstract concepts is enhanced in humans by the bigger brain size when compared with other animals [3]. In a more recent work, Zaidel [4] even traces the biological roots of art back to animal courtship displays, where the individual's fitness is paraded for inspection by potential partners.

Art is capable of influencing a viewer or listener's emotion and mood [5], and such ability has been explored a plethora of times in commercial advertisements and movies to engender different emotions from users [6]. In this work, we focus on the intimate relationship between visual arts and music [1]. We believe that further research in that area has the potential to positively affect human lives. For example, this relationship can be used to augment the perception ability of visually- and/or hearing-impaired people, allowing them to perceive the field of expression they are unable to. Another example is that with the rapid development of robotic technology in the field of engineering, humanoid robots are now being expected to effectively interact with humans, and the key to that is understanding a user's emotion by enabling the user with emotional intelligence.

With these goals in mind, Affective Computing researchers have delved into emotion prediction induced by visual stimuli for various applications [7, 8].

In a study, Kim [9] takes a video, which is assumed to have musical and visual components, and extracts its emotion indicators. Audio features, namely loudness, melody, and tempo, are extracted from the musical components, and visual features, namely orientation, hue, saturation, and intensity (or lightness), are extracted from the visual components. In addition to that, Kim obtains the valence and arousal indicators by manipulating EEG signals and uses the Adaptive Neuro-Fuzzy Inference System (ANFIS) as an emotion classifier. In a comparable research, Lee et al. [10] implements a model based on EEG and 3D fuzzy visual features that perceives a subjects' state of emotion by simultaneously analyzing a video clip and EEG signals from the subjects' brain while exposed to the same visual stimuli.

Research has also been conducted solely considering how visual stimuli affect a user's emotional state [11]. Yanulevskaya et al. [12] show that machines are capable of deriving emotion from paintings by extracting image texture features in the IAPS dataset and using a Support Vector Classifier to predict emotional scores. Zhang and Lee [13] build a machine with emotional features that considers EEG, image features, and subjects interaction to analyze images and learn more complex emotions. Lastly, Mikels et al. [14] uses images in the already-mentioned IAPS dataset to show that some images evoke one emotion more than others and that the dataset used is valuable when examining discrete emotions.

The aforementioned researchers have shown that visual stimuli can elicit emotion just as audio stimuli can. However, there is still a lot of work to be carried out with respect to the relationship between audio and video. In other words, it is utterly important to find a method that is able to bridge the gap between these two modalities of art. The following softwares aim to generate audio from an image: Photosounder [15], Paint2Sound [16], and SonicPhoto [17]. Photosounder is the "first audio editor/synthesizer to have an entirely image-based approach to sound editing and creation," and it can convert audio into images and vice-versa. Other mentioned applications are only able to convert images to sound and not the other way around. SonicPhoto is inspired by the first mentioned application, and even though it does not have all the features that Photosounder has, the creator claims that it has an automatic and convincing stereo and "a unique harmony filter to help create distinct and professional effects." Lastly, Paint2Sound generates audio by summing all synthesized sine waves from each pixel row assuming that each color of the image pixel represents a frequency band and the brightness of the pixels represent the amplitude. It is worth mentioning that the audio it generates is not very smooth or music-like as one would like.

All the previously mentioned applications are limited to the fact that they overlook the video's emotional characteristics in the music generation step. Zhao et al. [18] explore the task of emotion-based image musicalization by comparing the emotion obtained from visual and audio features and matching it with the best music in a pre-existing database. However, they only consider static images instead of videos and are unable to generate novel music due to their match and select approach, being limited to a given database. Moreover, current methods require a high level of human effort and skills in order to complete their task, whereas we believe that machine learning techniques are necessary in order to establish a richer relationship between audio and video.

Thus, the goal of this research is to convert visual information into its complementary emotionally charged audio using a deep neural network model that considers that important emotion aspect of scenes and music are alike. The proposed model uses important scene features such as its Hue, Saturation, and Intensity (HSI) components as input to an Adaptive Neuro-Fuzzy Inference System (ANFIS) [19] to classify its emotion aspect and thus choose one of the four Deep Recurrent Neural Networks (RNN) of the type Long Short-Term Memory (LSTM) [20]. ANFIS was chosen as a classifier because it is a hybrid inference system that combines the strengths of neural networks and fuzzy logic. Previous researchers [21] have shown that it is more reasonable to model emotion according to fuzzy rather than binary mathematics, and since we need to classify a set of features into their respective emotional classes, which contain vague information, the ANFIS network is the ideal model. As for the LSTM-RNN model, it is chosen to generate the sound features as our data, scene, and music are alike and have dynamic time properties that simply cannot be ignored. Other networks, such as feedforward neural networks, are not able to model dynamic data due to not having access to the cell's previous hidden state. This unique property enables RNNs to model sequence data well, and thus it was chosen to compose our model.

The proposed model considers the proposed visual features with ANFIS to successfully represent emotional features in a video. Once ANFIS is used to classify the visual features into emotional classes, each of the mentioned LSTMs is used to estimate the scene's variation through time represented by music features such as tempo, loudness, and rhythm given the image features. The RNN that will be used is determined by the classification result obtained from the ANFIS, which can have a positive or negative valence and a high or low arousal. The proposed method is then evaluated through a comparison of the target and estimated audio output spectrograms. The choice of using a spectrogram for comparison is due to it being a visual representation of the music and thus easy and intuitive to analyze. Quantitative evaluation is performed by calculating the Mean Absolute Error (MAE) between the original and generated music samples and by obtaining the Mean Opinion Scores (MOS) from human subjects. The generated music is also given as supplementary material (available here) for qualitative evaluation.

This paper is an extension of our previous work [22] with some major differences, such as the current model implements Deep LSTM models instead of shallow RNN models. The number of considered emotions is doubled by adding a second dimension, arousal, to the 1D valence emotion system, meaning that the ANFIS model is more complex, with 4 membership functions instead of 2, and our overall architecture is also more complex with 4 Deep LSTM models instead of just 2 shallow RNNs. The number of subjects providing MOS is also increased, along with more detailed explanations and diagrams regarding the algorithm. Furthermore, both the variance and mean of the MOS are being used for comparison, along with more samples added to the results, and a more in depth analysis of the dataset by including the video excerpts distribution and sample images from each class. Qualitative and quantitative evaluations are performed in order to effectively compare the current proposed model and our previous work. Qualitative evaluations include providing the generated music videos for the reader to access and form their own subjective opinions and pairwise comparison between samples, obtained from human subjects, to indicate model preference. Quantitative evaluation is conducted through a comparison of the MAE values, where lower values mean that the model is better able to perform domain transformation from visual to audio features.

The remainder of this paper is organized as follows. The coming section covers the methods and background information needed for our paper, including ANFIS, RNN, and emotion representation. This is followed by a section explaining the evaluation method and the proposed method, covering feature extraction, emotion classification, and music generation. We then introduce the dataset and the experimental results, which include discussion and comparison evaluations. Finally, we conclude this paper with the conclusion and future works.

## 2. Related Works

This section provides a brief explanation on the approaches used in our model and introduces the emotion representation used here.

*2.1. Adaptive Neuro-Fuzzy Inference System.* Adaptive Neuro-Fuzzy Inference System (ANFIS) [19] is a hybrid inference system that combines the strengths of neural networks and fuzzy logic. ANFIS is capable of approximating nonlinear functions and thus considered to be a universal estimator, by learning a set of fuzzy IF-THEN rules given a number of input-output pairs [23]. This is performed by mapping inputs to the fuzzy domain using input membership functions, in a process called fuzzification, computing the degree of membership of each input, generating rules, and mapping fuzzy outputs back into the probability domain through output membership functions, in a process called defuzzification.

The fuzzy inference system (FIS) used here is of Sugeno type [24], and it includes an adaptive learning algorithm that is able to identify the parameters and rules in the membership function. Consider the input vector $X = [x_1, x_2, \ldots, x_n]$ and the output vector $Y = [y_1, y_2, \ldots, y_n]$. The rule set for a first-order Sugeno fuzzy model is

$$\text{IF}(x_1 \text{ is } A_1^k)\text{AND}(x_2 \text{ is } A_2^k)\cdots(x_n \text{ is } A_n^k), \\ \text{THEN } y_k = f^k(x_1, x_2, \ldots, x_n). \quad (1)$$

The first part of each rule is defined as a fuzzy AND proposition where $A_j^k$ represents the $j$th variable defined by the membership function $\mu_j^k$. The following part of the rule is the first-order polynomial of the input vector $X$, which, when given, allows for a prediction of the variable values $Y$ [25].

The membership function employed here is of Bell type because it gives a slightly better performance than the Gaussian type membership function. The membership function is in the form of equation (2), where $c_j^k$ is the center of the curve, $a_j^k$ is the half width of the curve, and $b_j^k > 0$, which together with $a_j^k$, controls the slopes at the crossover points.

$$\mu_j^k(x_j) = \frac{1}{1 + \left|(x_j - c_j^k)/a_j^k\right|^{2b_j^k}}. \quad (2)$$

The degree of fulfillment for the $k$th rule is calculated using equation (3) and the inferred output can then be obtained by equation (4).

$$\mu^k(x) = \prod_{j=1}^{n} \mu_j^k(x_j), \quad k = 1, 2, \ldots, K. \quad (3)$$

$$y_i = \frac{\sum_{k=1}^{K} \mu^k(x) f^i}{\sum_{k=1}^{K} \mu^k(x)}. \quad (4)$$

The outcome is the construction of an adaptive network equivalent to a Sugeno fuzzy model.

*2.2. Recurrent Neural Networks.* Recurrent Neural Networks (RNNs) are able to model sequences by accessing their own previous hidden state in addition to the current input sample information [26]. The current hidden state $h_t$ is updated as follows:

$$h_t = \sigma(Wx_t + Uh_{t-1}). \quad (5)$$

where $x_t$ and $h_{t-1}$ are the current input and previous hidden states, respectively, and $W$ and $U$ are their corresponding weight matrices.

Limitations when using RNNs arise when dealing with longer sequences due to their vanishing and exploding gradients problem [27]. To remedy such issues, Hochreiter and Schmidhuber [20] introduce Long Short-Term Memory (LSTM) Recurrent Neural Networks. The authors propose a series of input $i_t$, output $o_t$, and forget $f_t$ gating mechanisms that allow the model to better control the flow of information. The incorporation of gates together with a memory cell $C_t$ mechanism that stores information for longer periods of time enables LSTM-RNNs to outperform vanilla RNNs. This model is mathematically formalized in equations (6a)–(6f), and as can be seen, the new proposed hidden state considers the output gate and the cell's current state instead of just the previous hidden state as in vanilla RNNs.

$$f_t = \sigma(W_{xf}x_t + W_{hf}h_{t-1}), \tag{6a}$$

$$i_t = \sigma(W_{xi}x_t + W_{hi}h_{t-1}), \tag{6b}$$

$$C_t = \tanh(W_{xC}x_t + W_{hC}h_{t-1}), \tag{6c}$$

$$C_t = f_t C_{t-1} + i_t C_t, \tag{6d}$$

$$o_t = \sigma(W_{xo}x_t + W_{ho}h_{t-1}), \tag{6e}$$

$$h_t = o_t \tanh(C_t). \tag{6f}$$

RNNs and its many variations have been widely used in tasks involving sequence learning [28, 29] and domain transformation [30, 31]. Given the dynamic nature of our task, it makes sense to explore the RNNs sequence-modeling abilities, specifically RNNs with LSTM cells that do not suffer from vanishing gradient problems. The use of LSTM-RNN also increases the depth of our model and improves its performance.

*2.3. Emotion Representation.* This research considers Lang's proposed two-dimensional scale [32], on which emotions are profiled according to their arousal and valence. That fixes our emotion domain in a 2D-axis, as shown in Figure 1, essentially meaning that our system deals with 4 emotions: positive low and high arousal and negative low and high arousal.

In order to obtain the emotion measurements for each video excerpt in the dataset, we conduct a survey with ten healthy subjects in their 20s and 30s. The volunteers are shown scene images and asked to rate their emotion as negative or positive and the strength of that emotion on a 9-point rating scale each, 1 meaning that the emotion is very negative or has low arousal and 9 being the score for a very positive emotion or one with high arousal characteristics. The average of the 10 scores is calculated and thus obtaining the MOS, which represents the emotion of a scene and which will be used as target in the supervised training of the ANFIS.

## 3. The Proposed Method

The proposed model is a novel hybrid deep neural network that uses an ANFIS to predict a video's emotion from its visual features and an LSTM-RNN to generate its corresponding audio features. The gathered audio features are then used to restore the audio original waveform and thus compose the entire audio corresponding to a scene with similar emotional characteristics. The novelty of our proposed method lies in the extraction of visual emotional features in order to transform them into audio signals with corresponding emotional aspects for users. Considering the proposed visual features with ANFIS, we can successfully represent emotional features in a video. Then, we propose a novel method to transform the obtained visual features into audio signals with similar emotional inkling using LSTM-RNN. The proposed combination results in a structure that can be used for other tasks involving modality transformation. Some direct practical applications include augmenting the perception ability of visually- and/or hearing-impaired people, implementing automatic movie directors, and assisting video content creators.

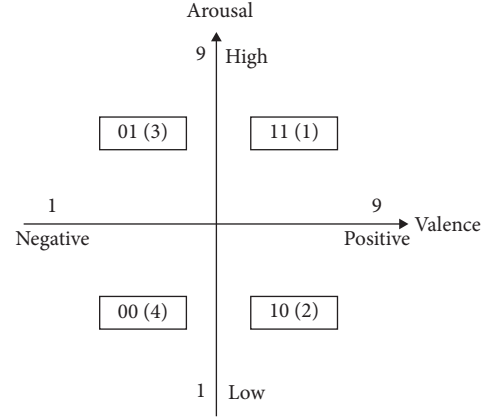

Figure 1: Two-dimensional valence-arousal emotion axis.

The proposed hybrid system's overall architecture for emotional video to audio transformation is shown in Figure 2. It comprises of three main modules: visual and audio feature extraction, emotion classification with ANFIS, and music generation with four LSTM-RNN. In this broad summary of the model, we explain the training and experimental stages individually and then provide a more detailed explanation of the aforementioned modules in the subsequent subsections.

*3.1. Training Stage.* The proposed model is trained in two parts: ANFIS training and LSTM-RNN training. During the ANFIS training stage, shown in Figure 2(a), the system receives as input a video without its sound component (a sequence of image frames) and extracts its visual information. It then performs clustering to construct emotion descriptors to give as input for the ANFIS to be trained using MOS as the target. After training, the system obtains knowledge about human emotions, divided into four groups, and is able to classify it accordingly. The videos given as input belong to one of the four groups of emotions considered here: positive/negative and high/low arousal. The second stage of training consists of training four RNNs in a similar way, so for simplicity, Figure 2(b) only shows the training of one of them. In this stage, the same video is given as input, but this time its sound component is also considered. Both the visual and audio features are extracted and given to the RNN responsible for the emotion group to which the video belongs to, to serve as input and target during training. In the end, we obtain four trained RNNs that are able to output sound features given visual features of similar emotion score.

*3.2. Experimental Stage.* In the experimental stage, shown in Figure 2(c), all trained models are used together as one. A video is given as input and its visual features are extracted and classified with the ANFIS into their respective MOS. This determines which of the four RNNs is responsible for generating the audio features relative to that excerpt which in turn allows for music to be generated.

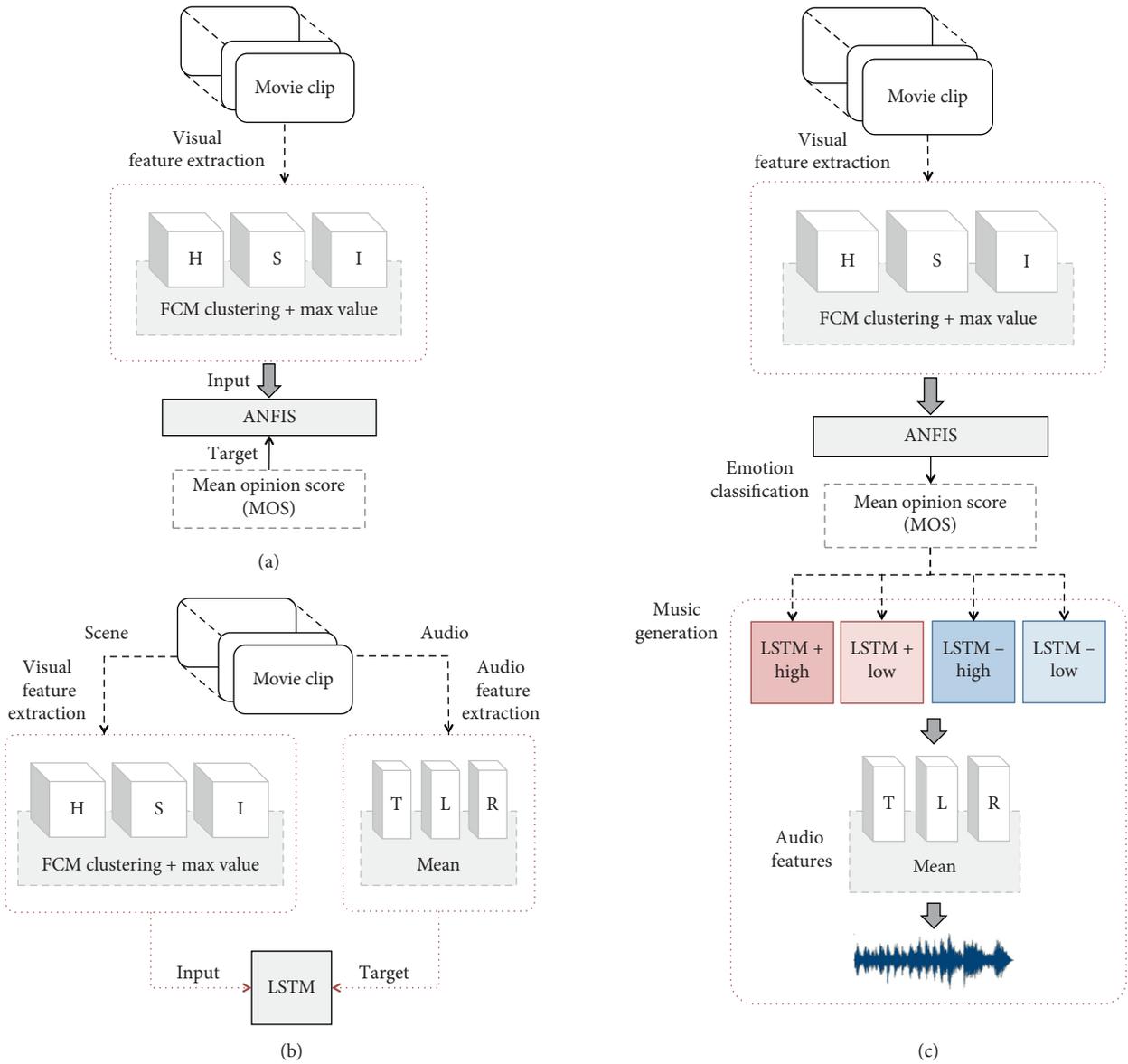

Figure 2: Overall architecture of the proposed method, with (a) ANFIS being trained for emotion classification from visual features, (b) deep LSTM-RNN being trained for domain transformation from visual to audio features, and (c) experimental stage for music generation from visual features.

### 3.3. Audio and Visual Feature Extraction.
In the feature extraction stage, the visual and audio features of a video containing scene and audio components are extracted, as shown in Figure 3. Our work uses a tweaked version of a study conducted by Kim [9] in order to obtain both audio and scene features, one modification being that it forfeits the use of the scene's orientation feature. Additionally, in Kim's original work, the audio features are obtained from the entire audio file, whereas in our work, the audio is segmented and features are extracted from each one of those splices while also storing the original equivalent audio segment for future use.

#### 3.3.1. Audio Feature Extraction.
In [9], three features are extracted from the audio: Tempo, Loudness, and Rhythm (TLR). *Tempo* is the underlying beat of the music and can also be thought of as the speed of a song [33]. It is obtained by gauging where the strongest autocorrelation occurs in the signal away from the origin [34]. This is achieved by first dividing the input audio signal into multiple splices of 0.1 s and resampling them to 8 kHz. Short Time Fourier Transform (STFT) is then used to obtain the Mel spectrogram (in decibels) of those signals via weighted summing of the spectrogram values. Then, the first-order difference along time is calculated and the positive differences are summed. The signal is then smoothed with a Gaussian envelope in order to get a one-dimensional onset envelope $O(t)$ as a function of time. The inner product is calculated between this onset envelope and its delayed version to obtain the autocorrelation, which shows the periodic structure of the envelope and which is used to check the peaks with large correlation. The next step is to apply a perceptual weighting window to the raw autocorrelation to simulate human tempo perception, which has a bias of 120 beats per minute (bpm). This step makes sure that peaks far from this bias are not as relevant as other peaks. The tempo period strength $\text{TPS}(\tau)$ is then calculated as follows:

$$\text{TPS}(\tau) = W(\tau) \sum_t O(\tau) O(t - \tau), \qquad (7)$$

where $W$ is a Gaussian weighting function on a log-time axis:

$$W(\tau) = \exp\left(-\left(\frac{1}{2}\right)\right)\left(\frac{(\log_2 (\tau/\tau_0))}{\sigma_\tau}\right)^2, \qquad (8)$$

where $\tau_0$ is the center of the tempo period bias and $\sigma_\tau$ controls the width of the weighting curve. Finally, *tempo* is estimated as the the largest $\text{TPS}(\tau)$.

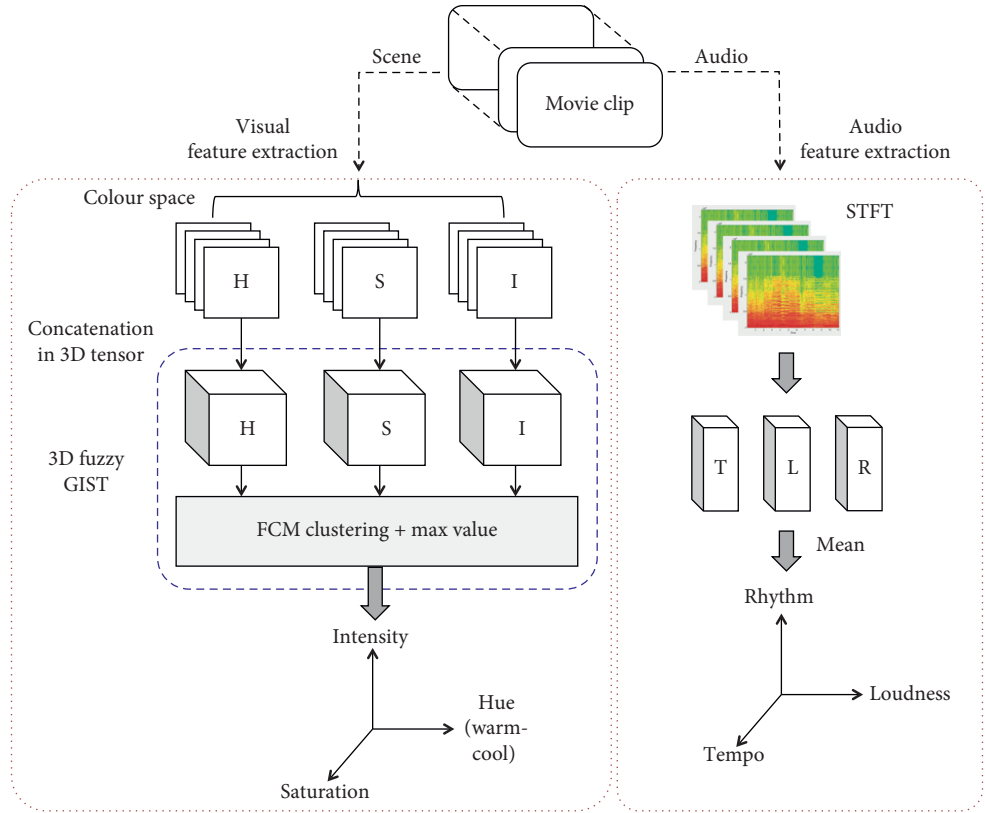

Figure 3: Visual and audio feature extraction stages.

*Loudness* is an acoustic term defined as the auditory sensation in which sound ranges from quiet to loud [35]. It is estimated by mapping the sound pressures in decibels as modeled in Figure 4. This is performed with a time-frequency decomposition that reflects the human ear response compensation. That means that loudness can be regarded as a personal subjective characteristic of sound or a direct psychological correlate of its magnitude, making it critical information. More weight is attributed to frequencies between 2 kHz and 5 kHz since that is where humans are most sensitive.

Lastly, *rhythm* is the way that particular sounds, including silence, of different durations are organized [33]. It is reflected in terms of tone information, which encompasses the frequency characteristics of the audio and the gradient of frequency variation over its period of time, parameters which have been tested and tapered by a manifold of previous experiments. The extraction method is shown in Figure 5, and it is obtained by calculating the log of the variance in the power spectrum density calculated with a fast Fourier transform.

*3.3.2. Visual Feature Extraction.* The visual feature extraction is performed by applying 3D fuzzy GIST [10] based on a tensor data of same dimension that includes the Hue (H)-Saturation (S)-Intensity (I) color space and the scene's dynamic properties in the third dimension. These tensor data are $M \times M \times T$, where $T$ is the duration, or more specifically the number of frames, of a scene, and $M \times M$ is the width and height of a single frame ($M = 256$). In order to effectively extract the visual information, a Fuzzy $c$-Means (FCM) [36] algorithm is used. The FCM algorithm can be thought of as a soft $k$-means algorithm [37]. In the $k$-means algorithm, the data are clustered into $k$ clusters, and a single sample can only belong to one cluster, whereas in the c-means algorithm, each input sample has a degree of belonging to each and every cluster, in true fuzzy manner. The FCM algorithm is used here to cluster the color information into three clusters. The outcome is that each frame has a $3 \times 1$ descriptor for each $H \times S \times I$ color feature, or, a final $9 \times 1$

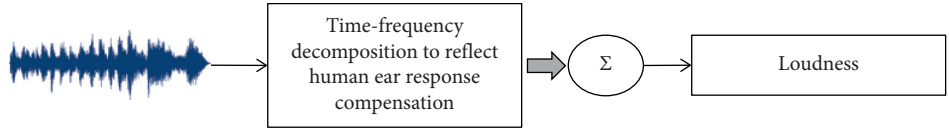

Figure 4: Loudness modeling procedure.

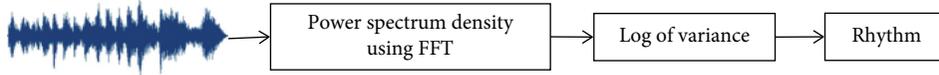

Figure 5: Rhythm feature extraction.

descriptor obtained by concatenating the mentioned $3 \times 1$ descriptors.

As a solution to simplify the model and to cut back the neural network's training time, we alter this feature extraction method. In this alteration, instead of simply concatenating the three $3 \times 1$ descriptors, the maximum value in each $3 \times 1$ descriptor is taken, obtaining three single values each representing their own color component. Those values are later concatenated, and the result is a $3 \times 1$ descriptor representing the Hue, Saturation, and Intensity information. In a physical sense, Hue is the attribute that allows humans to recognize a color as red, yellow, green, cyan, blue, or magenta [38]. Saturation represents the brilliance of a color, or how pure it is, and Intensity is a subjective term that indicates the brightness of a color.

### 3.4. Emotion Classification with ANFIS.

As previously mentioned and illustrated in Figure 1, there are four groups in which emotion can be classified: positive valence with high arousal, positive valence with low arousal, negative valence with high arousal, and negative valence with low arousal. Each scene can carry a different MOS valence and arousal emotion score and the network responsible for classifying the scene's visual features into one of those 4 groups is the ANFIS. A Fuzzy System is tasked with such assignment because it is more reasonable to model emotion according to fuzzy rather than binary mathematics [21]. The fuzzy inference system (FIS) used here is of the Sugeno type, and it includes an adaptive learning algorithm that is able to identify the parameters and rules in the membership function. Since it has been established that there are only 4 possible emotions, we use Matlab's *genfis* 1 function with 4 generalized Bell membership functions to generate the FIS parameters.

### 3.5. Music Generation with Deep LSTM-RNN.

A Long Short-Term Memory Recurrent Neural Network (LSTM-RNN) is chosen to estimate the sound features because our data, scene, and music are alike and have dynamic time properties that simply cannot be ignored. The RNN that will be used to estimate the output features is chosen according to the ANFIS scene's classification as negative or positive valence and low or high arousal.

The RNN used here has 2 hidden LSTM layers, making it a deep neural network, of 100 and 300 neurons, respectively, with the input being the scene features and the output being the sound features from which the resulting music can be obtained, as illustrated in Figure 6, and which should elicit emotions similar to those from the scene. The dataset used to train the RNNs is obtained by segmenting the original audios in the training video dataset and extracting their features. In the process of making such dataset, a dictionary of features plus equivalent audio is also stored to be used in the final stages of the model.

The initial step in the stage of music generation is obtaining each frame's respective audio features $d$ with the RNN chosen with the ANFIS classification. Afterward, we calculate the MAE of the difference between each and all of the features in the dataset used during training and the mentioned descriptor $d$. Equation (9) displays the MAE calculation of the difference between one set of features in the dataset $f$ and $d$:

$$\text{MAE} = \frac{1}{n} \sum_{j=1}^{n} |d_j - f_j|. \quad (9)$$

The following stage of music generation consists of applying the abovementioned equation to all stored features and selecting the one resulting in the smallest MAE to represent the frame in question. The final stage is to repeat this process with every frame in the scene and concatenate the generated audios such that there is only one audio representing the full scene in the end.

### 3.6. Evaluation Method.

This model is evaluated by presenting a segment of a scene, included during training, as input to the model so as to obtain the estimated audio output. The next step is to calculate the spectrogram of both audios and compare them to check for similarities. The spectrogram images will then allow users to make a visual comparison, and the MAE function will allow for a quantitative comparison. Similarity between two signals using MAE is decided depending on how small the value obtained is.

Further quantitative measure is obtained in the form of MOS from human evaluators on Amazon MTurk [39], an online marketplace for Human Intelligence Tasks called HITs. The emotion score includes the average μ, given in equation (11), and the variance Var, given in equation (10).

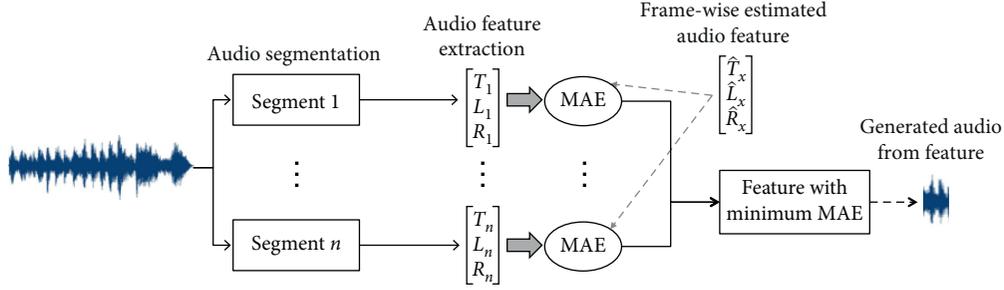

Figure 6: Music generation by obtaining audio segments referent to audio features.

$$\text{Var}(\widehat{Y}) = \frac{1}{N-1} \sum_{i=1}^{N} (\widehat{y}_i - \mu)^2. \quad (10)$$

$$\mu = \frac{1}{N} \sum_{i=1}^{N} \widehat{y}_i, \quad (11)$$

where $N$ is the number of emotion scores and $\widehat{Y} = [\widehat{y}_1, \widehat{y}_2, \ldots, \widehat{y}_n]$ is the vector with obtained emotion scores.

## 4. Experimental Results

*4.1. Dataset.* When searching for public datasets for our music generation from the video task, we are faced with the issue that there is currently no available dataset that is emotionally labeled and includes instrumental only-music videos. Realizing the need for such corpus, we take upon ourselves to create our own dataset in order to train the proposed model, which is composed of excerpts from eight instrumental music videos by Lindsey Stirling (Lindsey Stirling's online profile: https://www.youtube.com/user/lindseystomp), an American violinist, performance artist, dancer, composer, and singer. The music videos chosen can be accessed in a YouTube playlist (Dataset's YouTube playlist: https://www.youtube.com/playlist?list=PLg5IYs6I5_xPkTWQ6P_YOiTTh7IBlc7ZH). They are shown in Figure 7: Crystallize, Roundtable Rival, Beyond The Veil, Take Flight, Elements, Lord of the Rings Medley, Phantom of the Opera, and Moon Trance.

After selecting the music videos, we use crowd-sourcing to obtain their corresponding emotional labels. Table 1 shows the distribution of the video excerpts between the 4 considered emotions according to their valence-arousal scores: positive or negative valence and high or low arousal. It can be seen that the majority of the samples fall into the high arousal and positive valence group, followed by negative valence of same arousal and low arousal positive and negative following suit. Figure 8 complements the aforementioned table, by showing a sample image from each of the emotion classes.

*4.2. Results and Discussions.* First, we discuss the ANFIS classifier used to predict the emotion from a set of visual features obtained from an input video. The ANFIS classifier is tested with the training dataset and performs satisfactorily, with an accuracy of approximately 73.64% and learning curve as shown in Figure 9(a). We apply a dimensionality reduction technique, *t*-Distributed Stochastic Neighbor Embedding (*t*-SNE) [40], in Figures 9(b) and 9(c), to better visualize the target and estimated emotion scores for the three dimensional H × S × I input. In the aforementioned graphs, the cyan and blue dots indicate samples with negative low and high arousal emotion, respectively, and the red and green ones, positive high and low arousal, respectively. It can be observed in Figure 9(b) the occurrence of overlapping between samples of different classes, and as consequence, the classifier has some issues when trying to label the negative and positive samples that are too closely related to each other as well as when trying to differentiate between positive-high and low-arousal samples. It is highly likely that the reason for that is that those samples qualify as a neutral emotion and not as one of the extremes.

The quality of the model's generated audio is tested with two 6 seconds video segment from "Crystallize" and "Lord of the Rings Medley," see Figures 7(a) and Figures 7(f) (see Supplementary Materials Section). The spectrograms of the target outputs for those two examples are shown in Figures 10(a) and 10(c), and the spectrograms of the outputs estimated from our system are shown in Figures 10(b) and Figures 10(d), respectively. The MAE for the signals from the first video excerpt is calculated as 0.280, and the MOS collected is 6.43 ± 1.29 for valence and 6.57 ± 1.05 for arousal. As for the second video excerpt, the MAE for the signals is calculated as being 0.206, and the MOS collected is 5.86 ± 1.46 for valence and 4.71 ± 2.05 for arousal. The spectrograms show that, while the majority of the spectra are estimated with a reasonable accuracy, there are some gaps that fail to approximate. In order to mitigate this issue and achieve smoother transitions between audio segments, our music generating module can be improved by using a network that is able to model audio in its raw waveform [41, 42]. The drawback of modeling raw audio, however, is the increased input complexity and thus the increased model complexity. In order to prevent that, we instead build our model using higher level audio features, such as tempo, loudness, and rhythm, and higher level visual features, such as Hue, Saturation, and Intensity. Although this would greatly increase the complexity of the model, we cogitate that by modeling raw audio instead of mapping the audio features into pre-existing snippets, the generated music would sound less choppy and thus more pleasant to listen to, as it would seem to have more continuity.

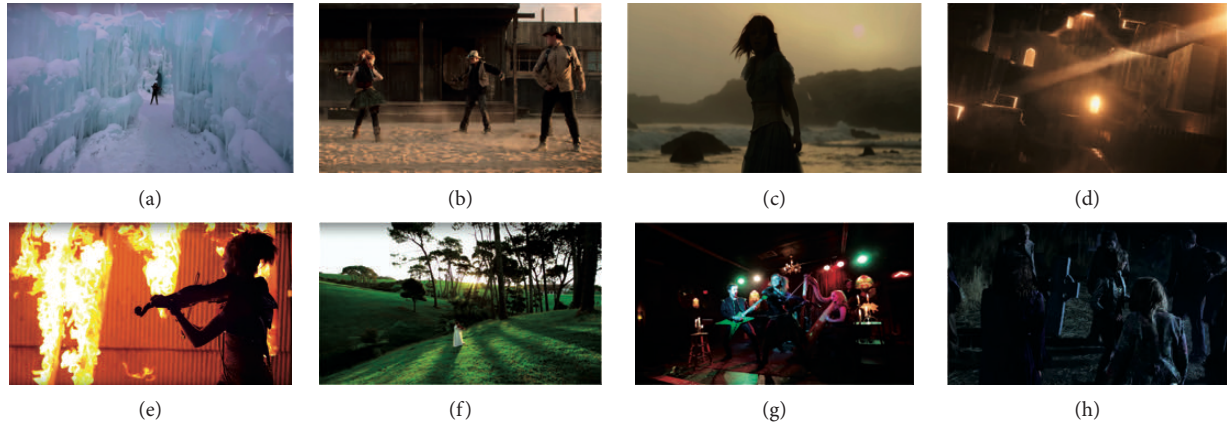

FIGURE 7: Lindsey Stirling Dataset: music video thumbnails for better visualization. (a) Crystallize, (b) Roundtable Rival, (c) Beyond The Veil, (d) Take Flight, (e) Elements, (f) Lord of the Rings Medley, (g) Phantom of the Opera, and (h) Moon Trance.

TABLE 1: Video excerpts distribution in the valence-arousal emotion domain for the Lindsey Stirling dataset.

| Arousal | Valence | | Total |
|---|---|---|---|
| | Positive | Negative | |
| High | 1,359 | 129 | 1,488 |
| Low | 70 | 47 | 117 |
| Total | 1,429 | 176 | 1,605 |

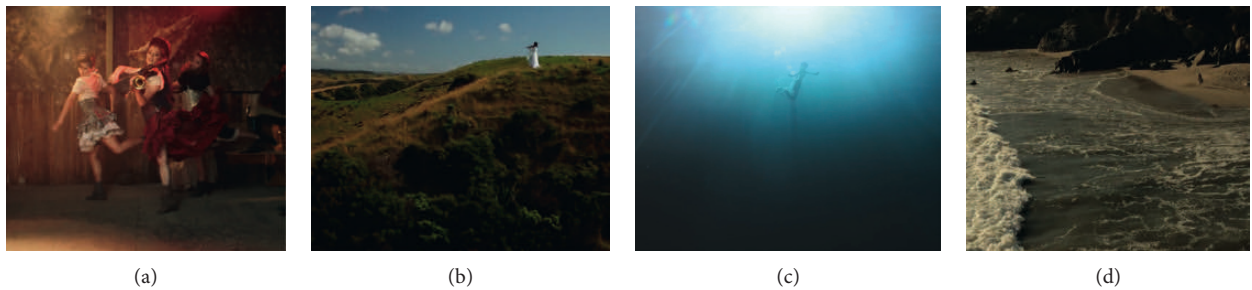

FIGURE 8: Sample images with different valence-arousal emotion tags from the dataset. (a) Positive-high. (b) Positive-low. (c) Negative-high. (d) Negative-low.

Further results are displayed in Table 2, which include the MAE and MOSs, with mean and variance, of eight 6 seconds sample scenes (music videos generated with our model can be found online at https://youtu.be/wNNkDTnyj4A) tested with our proposed model. Regarding the valence values, samples 2, 5, and 8 give back a reasonable result, while the rest tends to be in a more negative or neutral region of the emotion spectrum. The most likely explanation for the model not having a better performance is the nonconsideration of neutral emotion and the uneven distribution of data between the considered emotions as mentioned in Table 1. Regarding the arousal values, the scores indicate mild strength of emotions, meaning that even when the valence is correctly transmitted, the evoked emotion is not as strong or as soft as it should have been. Furthermore, the variance is not really significant in either the arousal or valence scores, meaning that the samples elicit similar emotion reactions from the subjects.

*4.2.1. Model Improvement Analysis.* In this section, we demonstrate the improvement in our model when compared with the previous work [22]. We evaluate both models on the same input videos and provide the generated music for qualitative evaluation (music videos generated with the previous model can be found online at https://youtu.be/7Gl8OvDjqp8). We use Amazon MTurk [39] to perform an unbiased pairwise comparison experiment on audios generated by both models. The experiment consists of showing an evaluator a number of paired music videos with the same visual stimuli but different music. The evaluator must then choose their preferred video from the options given, or "Either" if they do not have a particular preference. Our experiment, also referred to as an AB experiment, consists of 20 HITs, or twenty human evaluators, where the subjects are unaware of which model generated which music. Figure 11 shows the AB experiment with the eight provided generated samples. The color mapping is as follows: green for the current model, blue for the previous model, and orange for both. As can be seen, music generated by our model is preferred by evaluators more often than the ones generated by the previous model.

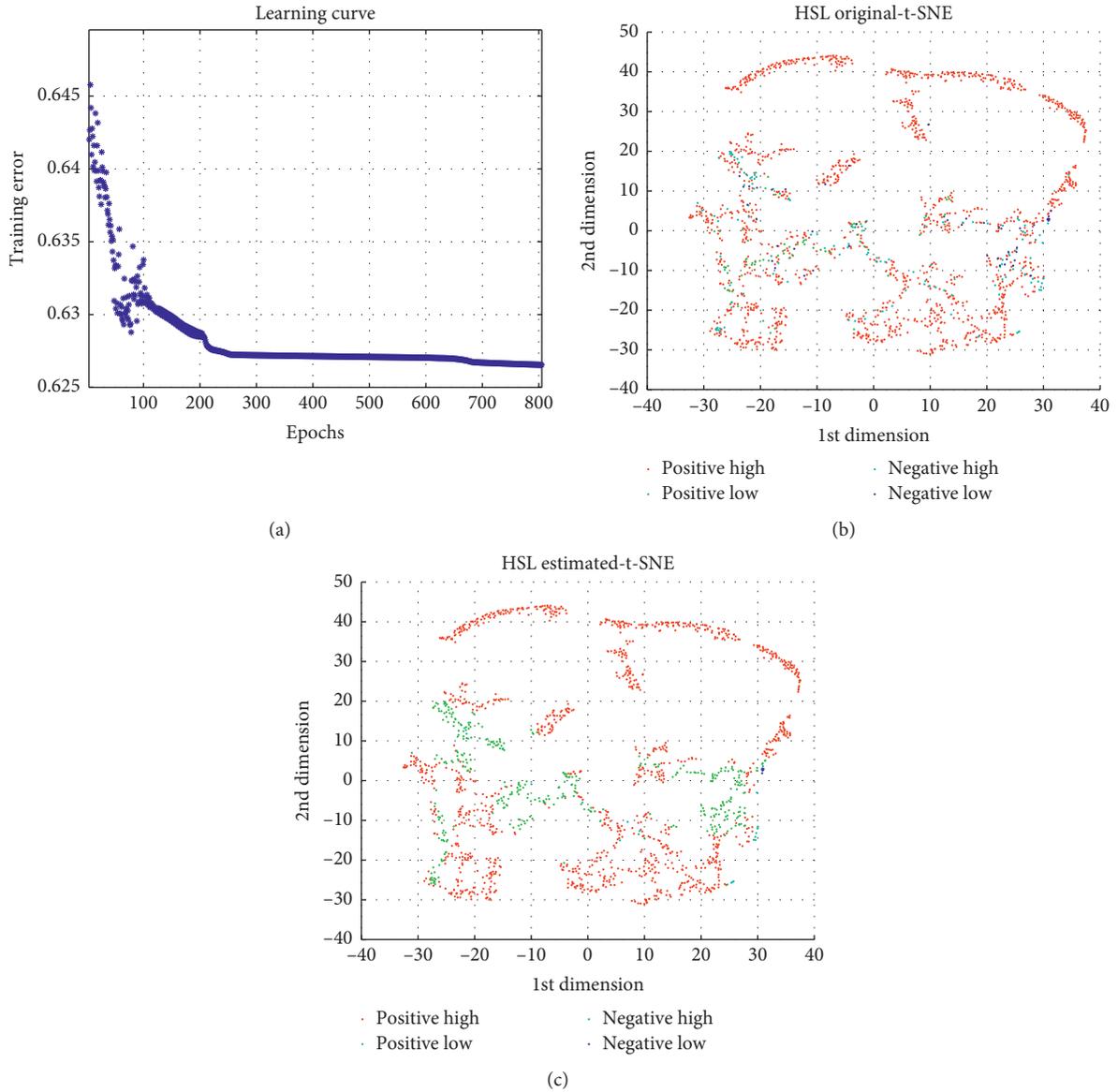

Figure 9: Results for ANFIS with H × S × I input with performance of ~73.64%, including (a) learning curve and (t)-SNE distribution for (b) target and (c) estimated emotion scores.

As a quantitative metric, we calculate the MAEs between each generated output and original audio and display them in Table 3, where the smaller the MAE, the better. As can be seen in the table, our model shows improvement of 6.47% in the MAE, with mean MAE of 0.217 against 0.232 in the previous model. These scores indicate that our model is able to perform domain transformation better than the previous model, and as a direct consequence, our model is better suited to recover appropriate music given a visual input.

After analyzing the qualitative and quantitative experimental results, we can conclude that our model shows better generative performance than the previous model. Furthermore, users show preference to music generated by our model than music generated by the baseline model.

*4.2.2. Evaluation on the DEAP Dataset.* Additional evaluation is performed on the DEAP dataset (https://www.eecs.qmul.ac.uk/mmv/datasets/deap) [43]. Although this dataset is not composed of instrumental music videos, it has most of the desired characteristics for our purpose. It is a collection of short 1-minute music videos annotated with emotion in a two-dimensional axis, meaning that it also contains valence and arousal labels. Figure 12 shows a few examples of music videos in this dataset, and Table 4 presents the data distribution between the considered emotion classes.

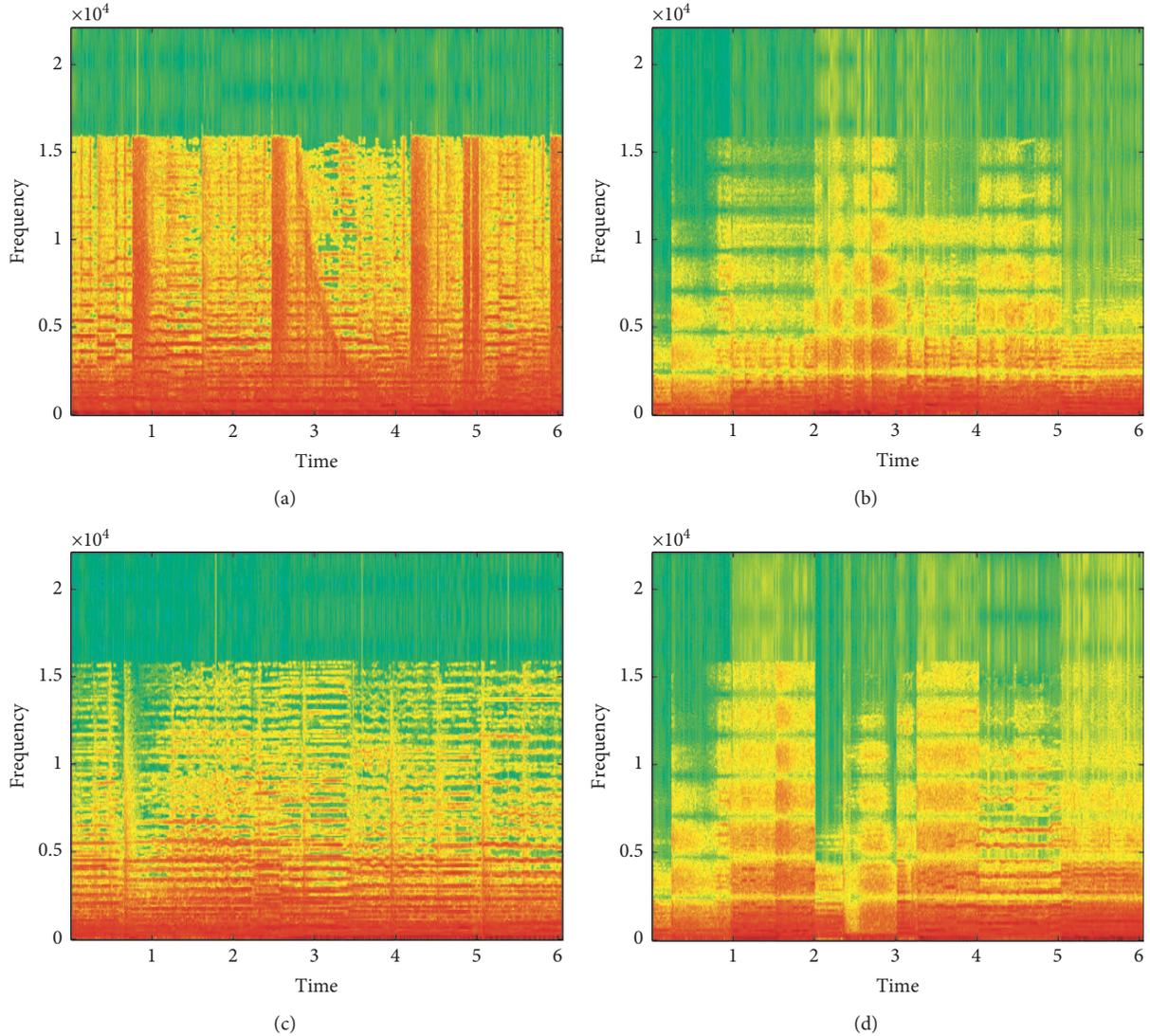

FIGURE 10: Spectrograms of the expected and estimated audio output (a, b) for *Crystallize* video excerpt, with MAE = 0.280 and MOS (Valence, Arousal) = (6.43 ± 1.29, 6.57 ± 1.05) and (c, d) for *LOTR* video excerpt, with MAE = 0.206 and MOS (valence, arousal) = (5.86 ± 1.46, 4.71 ± 2.05).

TABLE 2: Extended MOS and MAE results with 8 samples.

| # | MAE | Target MOS | | Obtained MOS | |
|---|---|---|---|---|---|
| | | Valence | Arousal | Valence | Arousal |
| 1 | 0.193 | 8.0 | 8.6 | 4.43 ± 1.99 | 5.14 ± 2.10 |
| 2 | 0.142 | 3.8 | 5.2 | 5.57 ± 1.84 | 5.57 ± 1.59 |
| 3 | 0.234 | 6.4 | 8.0 | 4.71 ± 1.58 | 5.00 ± 1.78 |
| 4 | 0.203 | 6.4 | 6.8 | 4.71 ± 1.39 | 3.71 ± 1.03 |
| 5 | 0.280 | 6.6 | 7.0 | 6.43 ± 1.29 | 6.57 ± 1.05 |
| 6 | 0.262 | 6.0 | 6.8 | 3.00 ± 0.53 | 4.29 ± 0.88 |
| 7 | 0.206 | 7.2 | 6.8 | 5.86 ± 1.46 | 4.71 ± 2.05 |
| 8 | 0.219 | 5.2 | 7.2 | 5.43 ± 1.18 | 5.86 ± 1.36 |

We evaluate our model with the DEAP dataset and provide ten samples of generated music videos (music videos generated with the DEAP dataset can be found at https://youtu.be/ZytjgqjjVRY). We also calculate the MAE to quantify the performance of our model in the domain transformation of visual inputs to audio outputs. Table 5 shows improvement of 4.49% in the MAE, with mean MAE of around 0.255 for our model against 0.267 for the previous model.

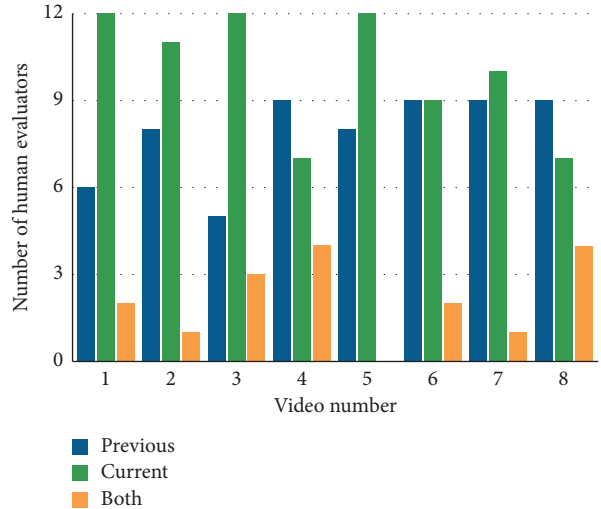

Figure 11: Pairwise comparison experiment between videos containing music generated by our model and the previous model. The color mapping is as follows: green (current model), blue (previous model), and orange (both).

Table 3: Comparison of MAE results calculated from music generated by our model and the previous one. Note that lower values are better.

| Video sample | 1 | 2 | 3 | 4 | 5 | 6 | 7 | 8 | Mean |
|---|---|---|---|---|---|---|---|---|---|
| Current | **0.193** | **0.142** | **0.234** | **0.203** | 0.280 | **0.262** | **0.206** | 0.219 | **0.217** |
| Previous [22] | 0.219 | 0.181 | 0.236 | 0.239 | **0.273** | 0.284 | 0.219 | **0.205** | 0.232 |

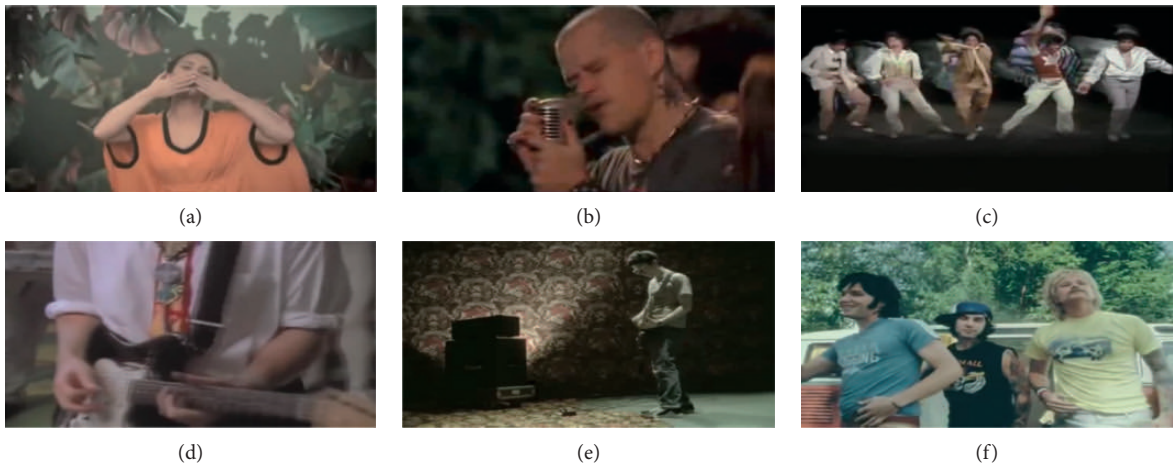

Figure 12: Examples of music videos in the DEAP dataset.

Analogous to the experiments on the Lindsey dataset, we also make use of Amazon MTurk [39] to obtain human evaluation on the emotional aspect of the generated music. Table 6 shows the MOS scores obtained from ten subjects. As can be seen, our model is able to retain the emotional characteristics from the original music video, with most samples scoring very similarly to the original emotion labels, except for samples 9 and 10 where the valence scores present slightly lower than the original values.

Table 4: Video excerpts distribution in the valence-arousal emotion domain for the DEAP dataset.

| | Valence | | |
|---|---|---|---|
| Arousal | Positive | Negative | Total |
| High | 250 | 350 | 600 |
| Low | 250 | 1,050 | 1,300 |
| Total | 500 | 1,400 | 1,900 |

Table 5: Comparison of MAE results calculated from music generated by our model and the previous one with the DEAP dataset. Note that lower values are better.

| Video sample | 1 | 2 | 3 | 4 | 5 | 6 | 7 | 8 | 9 | 10 | Mean |
|---|---|---|---|---|---|---|---|---|---|---|---|
| Current | **0.318** | **0.312** | 0.199 | **0.261** | 0.222 | **0.260** | 0.293 | **0.255** | 0.208 | 0.219 | **0.255** |
| Previous [22] | 0.329 | 0.318 | 0.228 | 0.264 | **0.203** | 0.281 | **0.262** | 0.290 | 0.214 | 0.284 | 0.267 |

Table 6: Human evaluation with 10 samples generated with the DEAP dataset using the current model.

| # | Target MOS | | Obtained MOS | |
|---|---|---|---|---|
|   | Valence | Arousal | Valence | Arousal |
| 1 | 5.39 | 5.32 | 5.40 ± 2.80 | 5.20 ± 2.91 |
| 2 | 5.58 | 5.18 | 4.10 ± 2.47 | 4.80 ± 2.74 |
| 3 | 5.02 | 5.29 | 5.00 ± 2.21 | 5.70 ± 2.50 |
| 4 | 5.53 | 5.55 | 6.00 ± 2.26 | 5.60 ± 2.84 |
| 5 | 5.01 | 5.44 | 4.60 ± 2.37 | 4.90 ± 2.69 |
| 6 | 5.49 | 4.20 | 4.40 ± 2.63 | 4.90 ± 2.81 |
| 7 | 5.30 | 5.04 | 5.40 ± 2.99 | 5.10 ± 3.18 |
| 8 | 4.72 | 5.50 | 4.70 ± 2.50 | 5.50 ± 2.51 |
| 9 | 5.13 | 5.21 | 3.60 ± 2.32 | 4.30 ± 3.16 |
| 10 | 5.11 | 4.92 | 3.80 ± 2.74 | 4.50 ± 3.27 |

The generated music, MAE, and human MOS scores indicate that our model is able to successfully perform multimodal domain transformation and generate audio with similar emotions to the visual inputs in the DEAP dataset.

## 5. Conclusion

In this work, we proposed a novel hybrid deep neural network that uses an ANFIS to predict a video's emotion from its visual features and an LSTM-RNN to generate audio features corresponding to the given visual features. The gathered audio features were then used to restore the audio original waveform and thus compose the entire audio corresponding to a scene with similar emotional characteristics. Our implementation's importance is due to the lack of a deep neuro-fuzzy model that is able to convert visual information into its respective audio features while also considering the important emotion aspect of scenes and music alike.

Our current work improved on the previous work by implementing deep LSTM models instead of shallow RNN models and considering double the number of emotions, meaning that the ANFIS model is more complex, with 4 membership functions instead of 2. The implemented model was evaluated qualitatively by providing the generated and target music videos as supplementary material and with *t*-SNE distribution comparison and quantitatively with MOS and MAE values. Improvements were shown by comparing our model with the ANFIS-Shallow RNN baseline model. Our results showed improvement of 6.47% in the MAE and similar global features in the spectrograms, indicating that our model was able to perform well in the domain transformation step between visual and audio features. Furthermore, human evaluation showed that the music generated by our model was able to elicit similar emotions from subjects in some of the samples, neutral in others and negative at times. We demonstrated that the relationship between scene and music could roughly be established by relating important features such as the Hue, Saturation, and Intensity of a scene to the tempo, loudness, and rhythm of a sound. However, the uneven distribution of data and the lack of a neutral emotion made it difficult for the model to perform better. Further qualitative and quantitative comparisons showed that our model improved on the previous model. Our model was able to improve the MAE by 4.49%, indicating that it was better at performing domain transformation from visual to audio features, and the music generated by our model was chosen more often than music generated by the previous model in a pairwise experiment performed on human evaluators. Our model's robustness was also evaluated on the DEAP dataset, also showing lower MAE values than the previous model and appropriate emotion scores when compared with the target MOS.

Future work will include an effort to balance the dataset more evenly among the four groups of emotions mentioned and take into consideration a neutral state on top of the already existing ones. We also aim to increase the size of our dataset. However, if making it robust proves to be too difficult, given the need to manually crawl the web and then obtain MOS scores from volunteers, we may also consider merging available instrumental music-only emotion datasets and videos emotion datasets according to their valence-arousal scores to make a new dataset. Furthermore, our model should be able to produce songs which are completely new and original instead of just comparing it to known audios. That could be achieved by including a Convolutional Auto-Encoder to generate new audio spectrograms or by exploring the possibility of using Generative Adversarial Networks.

## Data Availability

The dataset used to support the findings of this study are included within the article and is available online. Further information on the dataset is given in the Supplementary Materials.

## Conflicts of Interest

The authors declare that there are no conflicts of interest regarding the publication of this paper.

## Acknowledgments

This work was partly supported by Institute of Information and Communications Technology Planning & Evaluation

(IITP) grant funded by the Korea government (MSIT) (2016-0-00564, Development of Intelligent Interaction Technology Based on Context Awareness and Human Intention Understanding) and Korea Evaluation Institute of Industrial Technology (KEIT) grant funded by the Korea government (MOTIE) (50%) and the Technology Innovation Program: Industrial Strategic Technology Development Program (no: 10073162) funded by the Ministry of Trade, Industry & Energy (MOTIE, Korea) (50%).

## Supplementary Materials

The Supplementary Materials provided include experimental videos needed to supplement the demonstrated results. The files correspond to results mentioned in Figure 10 and Table 2, with the former consisting of two sets of generated and target music corresponding to the Crystallize and LOTR music videos. The latter are extended results, which also include the two videos from Figure 10, corresponding to eight test videos with generated music. The eight videos are available online (music videos for Table 2, generated with our model: https://youtu.be/wNNkDTnyj4A) and provided here together with the expected target music for comparison. The 8 music videos generated by the previous model are also available online (music videos generated with the previous model: https://youtu.be/7Gl8OvDjqp8). Additionally, the music generated with the DEAP dataset can be found online (https://youtu.be/ZytjgqjjVRY) and are provided together with the expected target music for comparison. (*Supplementary Materials*)